\documentclass[a4paper,12pt]{article}
\usepackage{amssymb}
\usepackage{amsmath}
\usepackage{epsfig}
\usepackage{stmaryrd}
\usepackage{graphics}

\usepackage{latexsym}
\usepackage{rotating}
\title{Traveling Wave Solutions of Degenerate Coupled Multi-KdV Equations}

\author {
Metin G\"{u}rses \thanks{Email:gurses@fen.bilkent.edu.tr} \ Asl{\i} Pekcan\thanks{Email:asli@fen.bilkent.edu.tr} \\
{\small Department of Mathematics, Science Faculty} \\
{\small Bilkent University,  06800 Ankara - Turkey}\\
{\small Department of Mathematics, Faculty of Science} \\
{\small Hacettepe University, 06800 Ankara - Turkey}}

\date{\nonumber}
\begin{document}

\maketitle
\date{\nonumber}

\baselineskip 17pt

\numberwithin{equation}{section}
\begin{abstract}
Traveling wave solutions of degenerate coupled $\ell$-KdV equations are studied. Due to symmetry reduction these equations
reduce to one ODE, $(f')^2=P_n(f)$ where $P_n(f)$ is a polynomial function of $f$ of degree $n=\ell+2$, where $\ell \geq 3$ in this work.
Here $\ell$ is the number of coupled fields.
There is no known  method to solve such ordinary differential equations when $\ell \geq 3$. For this purpose,
 we introduce two different type of methods to solve the reduced equation and
 apply these methods to degenerate three-coupled KdV equation.
One of the methods uses the Chebyshev's Theorem. In this case we find
several solutions some of which may correspond to solitary waves. The second method is a kind of factorizing the polynomial $P_n(f)$
as a product of lower degree polynomials. Each part of this product is assumed to satisfy different ODEs.\\

\textbf{Keywords:} Traveling wave solution, Degenerate KdV system, Chebyshev's theorem, Alternative method.
\end{abstract}

\newtheorem{thm}{Theorem}[section]
\newtheorem{Le}{Lemma}[section]
\newtheorem{defi}{Definition}[section]
\newtheorem{ex}{Example}[section]
\newtheorem{pro}{Proposition}[section]

\maketitle


\section{Introduction}

\noindent The system of degenerate coupled multi-field KdV equations is given as \cite{gur1}-\cite{anton3}
\begin{eqnarray}\label{eqn1}
\displaystyle u_t&=&\frac{3}{2}uu_x+q_x^2\nonumber \\
q_t^2&=&q^2u_x+\frac{1}{2}uq_x^2+q_x^3\nonumber \\
\vdots && \quad \vdots \quad \quad    \vdots \quad \quad \vdots\nonumber \\
q_t^{\ell-1}&=&q^{\ell-1}u_x+\frac{1}{2}uq_x^{\ell-1}+v_x \nonumber \\
v_t&=&-\frac{1}{4}u_{xxx}+v u_x+\frac{1}{2}uv_x,
\end{eqnarray}
where $q^1=u$ and $q^{\ell}=v$. In \cite{gur1}-\cite{gurnew}, it was shown that this system is also a degenerate KdV system of rank one.
In a previous work \cite{gur-pek} we focused
on the equation (\ref{eqn1}) for $\ell=2$. We reduced this equation into an ODE $(f')^2=P_4(f)$ where $P_4(f)$
is a polynomial function of degree four. We analyzed all possible cases about the zeros of $P_4(f)$. Due to this analysis we determined the cases
 when the solution is periodic or solitary. When the polynomial has one double $f_2$ and two simple zeros $f_1$, $f_3$ with $f_1< f_2< f_3$, or one triple and one simple zeros, the
 solution is solitary. Other cases give periodic or non-real solutions. By using the Jacobi elliptic functions \cite{Jacobi}, we obtained periodic solutions and all solitary wave
solutions which rapidly decay to some constants, explicitly. We have also shown that there are no real asymptotically vanishing traveling wave solutions for $\ell=2$. Indeed we have the following theorem for the degenerate coupled $\ell$-KdV equation, $\ell\geq 2$. The degenerate coupled $\ell$-KdV equation can be reduced to the equation
\begin{equation}\label{genpoly}
(f')^2=P_{\ell+2}(f)
\end{equation}
 by taking $\ell$
functions as $u(x,t)=q^1(x,t)=f(\xi)$, $q^2(x,t)=f_2(\xi)$, \ldots, $v(x,t)=q^{\ell}=f_{\ell}(\xi)$, where $\xi=x-ct$ in (\ref{eqn1}). Here
$P_{\ell+2}(f)$ is a polynomial of $f$ of degree $\ell+2$. If we apply the asymptotically vanishing
boundary conditions to (\ref{genpoly}), we have
 \begin{equation}
(f')^2=B f^2(f+2c)^{\ell}.
 \end{equation}
\begin{thm}\label{thmrealtravel} When $\ell=$ odd, we have $B > 0$, and then the degenerate coupled $\ell$-KdV equation has real traveling wave solution with asymptotically vanishing boundary conditions, but when $\ell=$ even, the constant $B < 0$. Hence the equation does not have a real traveling wave solution with asymptotically vanishing boundary conditions.
\end{thm}
\noindent \textbf{Proof.} Consider the degenerate coupled $\ell$-KdV equation (\ref{eqn1}). Let $u(x,t)=q^1(x,t)=f(\xi)$, $q^2(x,t)=f_2(\xi)$, \ldots, $v(x,t)=q^{\ell}=f_{\ell}(\xi)$ where $\xi=x-ct$. By using the first $\ell-1$ equations we obtain all functions $f_i(\xi)$, $i=2, 3, \ldots, \ell$ as a polynomial of $f(\xi)$. We get
\begin{eqnarray*}
f_2(\xi)&=&Q_2(f)=-cf-\frac{3}{4}f^2+d_1=-\alpha_2 f^2+A_1(f)\\
f_3(\xi)&=&Q_3(f)=\frac{3}{2}cf^2+\frac{1}{2}f^3+(c^2-d_1)f+d_2\\
&=&\alpha_3 f^3+A_2(f)\\
f_4(\xi)&=&Q_4(f)=-\frac{5}{16}f^4-\frac{3}{2}cf^3+\Big(-\frac{9}{4}c^2+\frac{3}{4}d_1\Big)f^2+(-c^3+cd_1-d_2)f+d_3\\
&=&\alpha_4 f^4+A_3(f)\\
&\vdots&\\
f_{\ell}(\xi)&=& Q_{\ell}(f)=(-1)^{\ell+1}\alpha_{\ell}f^{\ell}+A_{\ell-1}(f),
\end{eqnarray*}
where $\alpha_j>0$ are constants, $A_i(f)$ and $Q_j(f)$ are polynomials of $f$ of degree $i$ and $j$, $i=1, 2, \ldots, \ell-1$, $j=2, 3, \ldots, \ell$, respectively.

Now use the $\ell$th equation. We obtain
\begin{equation}
\displaystyle \frac{1}{4} f'''=(-1)^{\ell-1}\alpha_{\ell}f'f^{\ell}\Big(1+\frac{\ell}{2} \Big)+c(-1)^{\ell-1}\ell\alpha_{\ell}f'f^{\ell-1}+f'A_{\ell-1}+\frac{1}{2}f\frac{\partial A_{\ell-1}(f)}{\partial x}-\frac{\partial A_{\ell-1}(f)}{\partial t}.
\end{equation}
Integrating above equation once we get
$$
\displaystyle \frac{1}{4}f''=(-1)^{\ell-1}\frac{\alpha_{\ell}}{\ell+1}f^{\ell+1}\Big(1+\frac{\ell}{2} \Big)+R_{\ell}(f).
$$
By using $f'$ as an integrating factor, we integrate once more. Finally, we obtain
\begin{equation}
(f')^2=(-1)^{\ell-1}\frac{8\alpha_{\ell}}{(\ell+1)(\ell+2)}f^{\ell+2}\Big(1+\frac{\ell}{2} \Big)+R_{\ell+1}(f)=Bf^{\ell+2}+R_{\ell+1}(f),
\end{equation}
where $R_i(f)$ is a polynomial of $f$ of degree $i$, $i=\ell, \ell+1$. When $\ell$ is odd, the coefficient of $f^{\ell+2}$, that is $B$, is positive, and when
$\ell$ is even, $B$ is negative. Applying asymptotically vanishing boundary conditions, we get $(f')^2=B f^2(f+2c)^{\ell}$, where $\mathrm{sign}(B)=(-1)^{\ell-1}$. Hence for $\ell=$ odd, the degenerate coupled $\ell$-KdV equation has real traveling wave solution with asymptotically vanishing boundary conditions, but when $\ell=$ even, it does not. $\Box$

In this work, we study the equation (\ref{eqn1}) for $\ell=3$ which is
\begin{eqnarray}
\displaystyle u_t&=&\frac{3}{2}uu_x+v_x \label{l=3a} \\
v_t&=&vu_x+\frac{1}{2}uv_x+\omega_x \label{l=3b}\\
\omega_t&=&-\frac{1}{4}u_{xxx}+\omega u_x+\frac{1}{2}u\omega_x, \label{l=3c}
\end{eqnarray}
in detail. It is clear from Theorem \ref{thmrealtravel} that unlike the case $\ell=2$,
we have real traveling wave solution with asymptotically
vanishing boundary conditions in $\ell=3$ case.

\noindent Here the system (\ref{l=3a})-(\ref{l=3c}) reduces to a polynomial of degree five,
\begin{equation}\label{P_5(f)}
\displaystyle (f')^2=\frac{f^5}{2}+3cf^4+(6c^2-2d_1)f^3+4(c^3-cd_1+d_2)f^2+8d_3f+8d_4=P_5(f),
\end{equation}
where $f(\xi)=u(x,t)$, and $c, d_1, d_2, d_3, d_4$ are constants. When the degree of the polynomial in the reduced equation is equal to five or greater it is almost impossible to solve them. As far as we know there is no known method to solve these equations. We shall introduce two methods to solve such equations. The first one is based on the Chebyshev's Theorem \cite{Cheb} which is used recently to solve the Einstein field equations for a cosmological model \cite{CGLY}, \cite{CGY}. By using the Chebyshev's theorem we give several solutions of the reduced equations for $\ell=3$ and also for arbitrary $\ell$. The second method is based on the factorizing the polynomial $P_{\ell+2}(f)$ as product of lower degree polynomials. In this way we make use of the reduced equations of lower degrees. We have given all possible such solutions for $\ell=3$.

The layout of our paper is as follows: In Sec. II, we study the behavior of the solutions in the
neighborhood of the zeros of $P_5(f)$ and discuss the cases giving solitary wave solutions. In Sec. III, we find exact solutions of the system (\ref{l=3a})-(\ref{l=3c})
by a new method proposed for any $\ell\geq 3$ which uses the Chebyshev's Theorem and analyze the cases in which we may have solitary wave and kink-type solutions. In Sec. IV, we present an alternative method. Particularly, we find
 the solutions of the system (\ref{l=3a})-(\ref{l=3c}). Here we obtain many solutions for $\ell=3$ including solitary wave, kink-type, periodic, and unbounded solutions. We present some of them in the text but the rest of the solutions are given in the Appendices A and B.

\section{General Waves of Permanent Form for $(\ell=3)$}

\bigskip

\subsection{Zeros of $P_5(f)$ and Types of Solutions}

Here we will analyze the zeros of $P_5(f)$ in (\ref{P_5(f)}).

\noindent \textbf{(i)} \, If $f_1=f(\xi_1)$ is a \textit{simple zero} of $P_5(f)$ we have $P_5(f_1)=0$. Taylor
expansion of $P_5(f)$ about $f_1$ gives
\begin{eqnarray*}
(f')^2&=&P_5(f_1)+P_5'(f_1)(f-f_1)+O((f-f_1)^2)\\
&=&P_5'(f_1)(f-f_1)+O((f-f_1)^2).
\end{eqnarray*}
From here we get $f'(\xi_1)=0$ and $\displaystyle f''(\xi_1)=P_5'(f_1)/2$. Hence we can write the function $f(\xi)$ as
\begin{eqnarray}
\displaystyle f(\xi)&=&f(\xi_1)+(\xi-\xi_1)f'(\xi_1)+\frac{1}{2}(\xi-\xi_1)^2f''(\xi_1)+O((\xi-\xi_1)^3)\nonumber\\
&=& f_1+\frac{1}{4}(\xi-\xi_1)^2P_5'(f_1)+O((\xi-\xi_1)^3).
\end{eqnarray}
Thus, in the neighborhood of $\xi=\xi_1$, the function $f(\xi)$ has local minimum or maximum as $P_5'(f_1)$ is positive or negative respectively since
$\displaystyle f''(\xi_1)=P_5'(f_1)/2$.\\
\bigskip

\noindent \textbf{(ii)} \, If $f_1=f(\xi_1)$ is a \textit{double zero} of $P_5(f)$ we have $P_5(f_1)=P_5'(f_1)=0$. Taylor
expansion of $P_5(f)$ about $f_1$ gives
\begin{eqnarray}\label{doublezero}
\displaystyle (f')^2&=&P_5(f_1)+P_5'(f_1)(f-f_1)+\frac{1}{2}(f-f_1)^2P_5''(f_1)+O((f-f_1)^3)\nonumber \\
&=&\frac{1}{2}(f-f_1)^2P_5''(f_1)+O((f-f_1)^3).
\end{eqnarray}
To have real solution $f$, we should have $P_5''(f_1)> 0$. From the equality (\ref{doublezero}) we get
$$
\displaystyle f'\pm \frac{1}{\sqrt{2}}f\sqrt{P_5''(f_1)}\sim \pm \frac{1}{\sqrt{2}}f_1\sqrt{P_5''(f_1)},$$
which gives
\begin{equation}
\displaystyle f(\xi) \sim  f_1+\alpha e^{\pm \frac{1}{\sqrt{2}}\sqrt{P_5''(f_1)}\xi},
\end{equation}
where $\alpha$ is a constant. Hence $f\rightarrow f_1$ as $\xi\rightarrow \mp \infty$. The solution $f$ can have only one peak and the wave extends
from $-\infty$ to $\infty$.\\
\bigskip

\noindent \textbf{(iii)} \, If $f_1=f(\xi_1)$ is a \textit{triple zero} of $P_5(f)$ we have $P_5(f_1)=P_5'(f_1)=P_5''(f_1)=0$. Taylor
expansion of $P_5(f)$ about $f_1$ gives
\begin{eqnarray}\label{triplezero}
\displaystyle (f')^2&=&P_5(f_1)+P_5'(f_1)(f-f_1)+\frac{1}{2}(f-f_1)^2P_5''(f_1)+\frac{1}{6}(f-f_1)^3+O((f-f_1)^4)\nonumber \\
&=&\frac{1}{6}(f-f_1)^3P_5'''(f_1)+O((f-f_1)^4).
\end{eqnarray}
This is valid only if both signs of $(f-f_1)^3$ and $P_5'''(f_1)$ are same. Hence, to obtain real solution $f$ we have the following two possibilities:

\noindent $1)$ $(f-f_1)> 0$ and $P_5'''(f_1)>0,$ \\

\noindent $2)$ $(f-f_1)< 0$ and $P_5'''(f_1)<0.$\\

\noindent If $(f-f_1)> 0$ and $P_5'''(f_1)>0$ then we have
$$
\displaystyle f'\sim \pm \frac{1}{\sqrt{6}}(f-f_1)^{3/2}\sqrt{P_5'''(f_1)},$$
which gives
\begin{equation}
\displaystyle f(\xi)\sim f_1+\frac{4}{\Big(\pm \frac{1}{\sqrt{6}}\sqrt{P_5'''(f_1)}\xi+\alpha_1\Big)^2},
\end{equation}
where $\alpha_1$ is a constant. Thus $f\rightarrow f_1$ as $\xi\rightarrow \pm \infty$.

\noindent Let $(f-f_1)< 0$ and $P_5'''(f_1)<0$. Then
$$
\displaystyle f'\sim \pm \frac{1}{\sqrt{6}}(f_1-f)^{3/2}\sqrt{-P_5'''(f_1)},$$
which yields
\begin{equation}
\displaystyle f(\xi)\sim f_1-\frac{4}{\Big(\pm \frac{1}{\sqrt{6}}\sqrt{-P_5'''(f_1)}\xi+\alpha_2\Big)^2},
\end{equation}
where $\alpha_2$ is a constant. Thus $f\rightarrow f_1$ as $\xi\rightarrow \pm \infty$.

\bigskip

\noindent \textbf{(iv)} \,If $f_1=f(\xi_1)$ is a \textit{quadruple zero} of $P_5(f)$, then  we have $P_5(f_1)=P_5'(f_1)=P_5''(f_1)=P_5'''(f_1)=0$.
In this case Taylor expansion of $P_5(f)$ about $f_1$ gives
\begin{equation}\label{quadruplezero}
\displaystyle (f')^2=\frac{1}{24}(f-f_1)^4P_5^{(4)}(f_1)+O((f-f_1)^5).
\end{equation}
This is valid only if $P_5^{(4)}(f_1)>0$. Then we have
$$
\displaystyle f'\sim \pm \frac{1}{2\sqrt{6}}(f-f_1)^{2}\sqrt{P_5^{(4)}(f_1)},$$
which gives
\begin{equation}
\displaystyle f(\xi)\sim f_1-\frac{1}{ \pm \frac{1}{2\sqrt{6}}\sqrt{P_5^{(4)}(f_1)}\xi+\gamma_1},
\end{equation}
where $\gamma_1$ is a constant. Thus $f\rightarrow f_1$ as $\xi\rightarrow \pm \infty$.

\noindent \textbf{(v)} \,If $f_1=f(\xi_1)$ is a \textit{zero of multiplicity 5} of $P_5(f)$, then  we have $P_5(f)=(f-f_1)^5/2$.
This is valid only if $f-f_1> 0$. So we obtain the solution $f$ as
\begin{equation}\label{solutionmultfive}
\displaystyle f=f_1+\Big(\frac{4}{9}\Big)^{1/3}\frac{1}{\Big(\pm \frac{\xi}{\sqrt{2}}+m_1\Big)^{2/3}},
\end{equation}
where $m_1$ is a constant. Hence $f\rightarrow f_1$ as $\xi \rightarrow \pm \infty$.

\subsection{All Possible Cases Giving Solitary Wave Solutions}
We analyze all possible cases about the zeros of $P_5(f)$ that may give solitary wave solutions. Here in each cases we will present
the sketches of the graphs of $P_5(f)$. Real solutions of $(f')^2=P_5(f)\geq 0$ occur in the shaded regions.

\bigskip
\noindent \textbf{(1)\, One double and three simple zeros.}

\begin{figure}[!h]
\centering
\begin{center}
\includegraphics[angle=0,scale=.25]{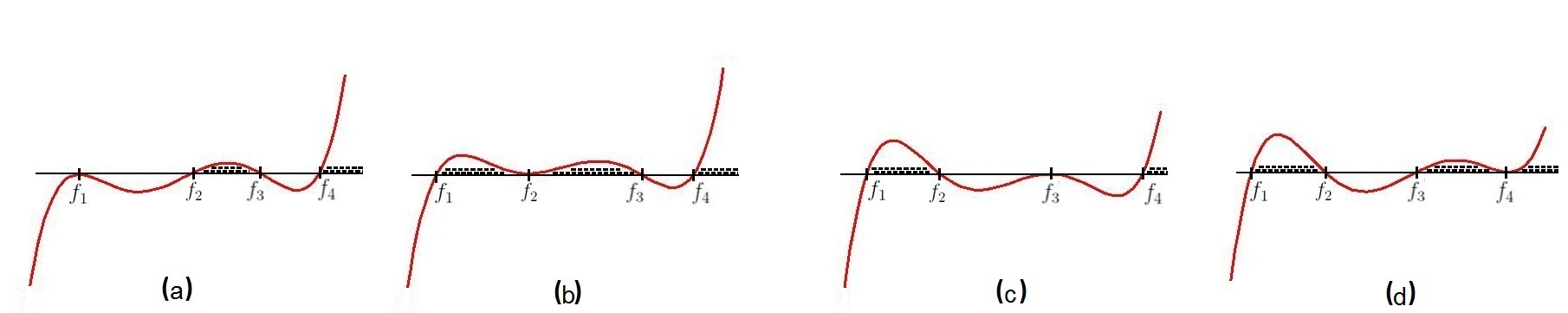}
\end{center}
\caption{Graphs of $P_5(f)$ having one double and three simple zeros}
\end{figure}

\noindent In Figure 1.\textbf{(b)}, $f_1$, $f_3$, and $f_4$ are simple zeros and $f_2$ is a double zero. The real solution occurs when $f$ stays between
$f_1$ and $f_2$ or $f_2$ and $f_3$. At $f_1$, $P_5'(f_1)=f''(\xi_1)> 0$ hence graph of the function $f$ is concave up at $\xi_1$. At double zero $f_2$,
$f\rightarrow f_2$ as $\xi \rightarrow \pm \infty$. Hence we have a solitary wave solution with amplitude $f_1-f_2<0$. Similarly at $f_3$, $P_5'(f_3)=f''(\xi_3)<0$, hence graph of the function $f$ is concave down at $\xi_3$. Therefore, we also have a solitary wave solution with amplitude $f_3-f_2>0$.

\noindent Now consider the graph \textbf{(d)} in Figure 1. For $f_3$ we have $P_5'(f_3)=f''(\xi_3)> 0$ thus graph of the function is concave up at $\xi_3$. At double zero $f_4$, $f\rightarrow f_4$ as $\xi \rightarrow \pm \infty$. Hence we have a solitary wave solution with amplitude $f_3-f_4<0$. In other cases we have periodic solutions.

\bigskip
\noindent \textbf{(2)\, Two double and one simple zeros.}

\begin{figure}[!h]
\centering
\begin{center}
\includegraphics[angle=0,scale=.25]{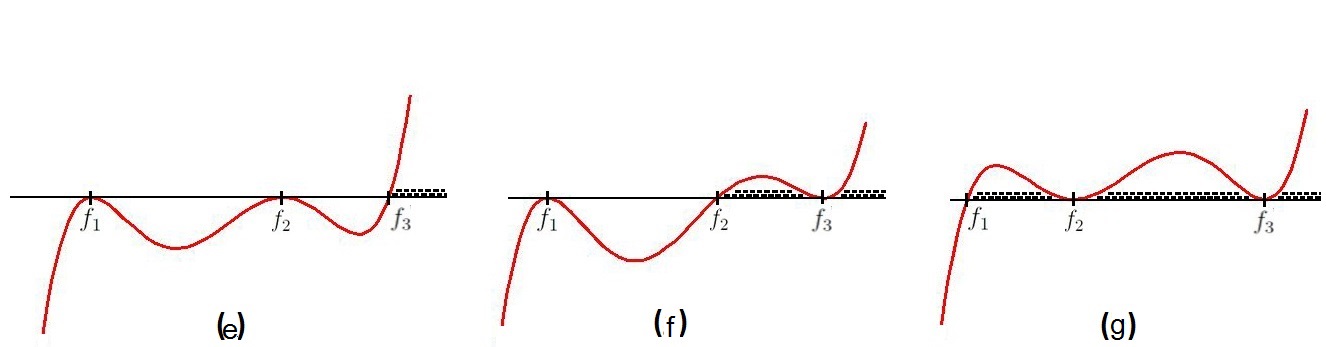}
\end{center}
\caption{Graphs of $P_5(f)$ having two double and one simple zeros}
\end{figure}

\noindent In Figure 2.\textbf{(f)}, $f_1$ and $f_3$ are double zeros and $f_2$ is a simple zero.  The real solution occurs when $f$ stays between
$f_2$ and $f_3$. For $f_2$ we have $P_5'(f_2)=f''(\xi_2)>0$ thus graph of the function is concave up at $\xi_2$. At double zero $f_3$, $f\rightarrow f_3$ as $\xi \rightarrow \pm \infty$. Hence we have a solitary wave solution with amplitude $f_2-f_3<0$.

\noindent Now consider the graph \textbf{(g)} in Figure 2. Here $f_2$ and $f_3$ are double zeros and $f_1$ is a simple zero.  The real solution occurs when $f$ stays between
$f_1$ and $f_2$ or $f_2$ and $f_3$. For $f_1$ we have $P_5'(f_1)=f''(\xi_1)>0$ thus graph of the function is concave up at $\xi_1$. At double zero $f_2$, $f\rightarrow f_2$ as $\xi \rightarrow \pm \infty$. Hence we have a solitary wave solution with amplitude $f_1-f_2<0$. The other cases give kink, anti-kink type or unbounded solutions.

\bigskip

\noindent \textbf{(3)\, One triple and two simple zeros.}

\begin{figure}[!h]
\centering
\begin{center}
\includegraphics[angle=0,scale=.25]{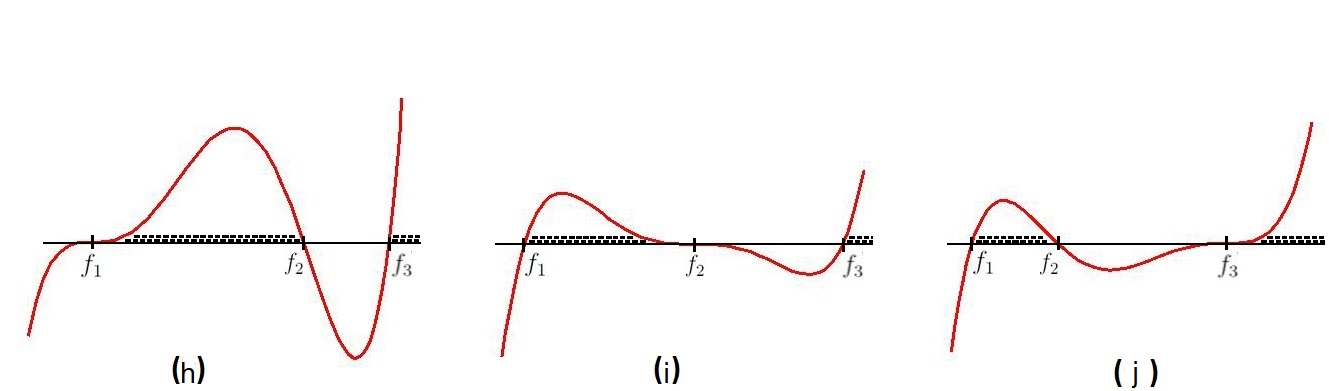}
\end{center}
\caption{Graphs of $P_5(f)$ having one triple and two simple zeros}
\end{figure}

\noindent Consider the graph \textbf{(h)} in Figure 3. Here $f_2$ and $f_3$ are simple zeros and $f_1$ is a triple zero. The real solution occurs when $f$ stays between
$f_1$ and $f_2$. For $f_2$ we have $P_5'(f_2)=f''(\xi_2)< 0$ so graph of the function $f$ is concave down at $\xi_2$. At triple zero $f_1$, $f\rightarrow f_1$ as $\xi \rightarrow \pm \infty$. Hence we may have a solitary wave solution with amplitude $f_2-f_1>0$.

\noindent In the graph in Figure 3.\textbf{(i)}, $f_1$ and $f_3$ are simple zeros and $f_2$ is a triple zero. The real solution occurs when $f$ stays between
$f_1$ and $f_2$. For $f_1$ we have $P_5'(f_1)=f''(\xi_1)> 0$ so graph of the function $f$ is concave up at $\xi_1$. At triple zero $f_2$, $f\rightarrow f_2$ as $\xi \rightarrow \pm \infty$. Hence we may have a solitary wave solution with amplitude $f_1-f_2< 0$. The other case gives periodic solution.

\newpage

\noindent \textbf{(4)\, One quadruple and one simple zeros.}

\begin{figure}[!h]
\centering
\begin{center}
\includegraphics[angle=0,scale=.25]{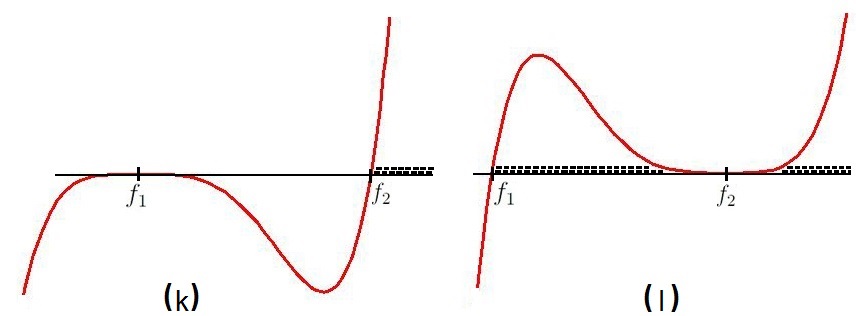}
\end{center}
\caption{Graphs of $P_5(f)$ having one quadruple and one simple zeros}
\end{figure}

\noindent In the graph Figure 4.\textbf{(l)}, $f_1$ is a simple zero and $f_2$ is a quadruple zero. The real solution occurs between $f_1$ and $f_2$. At $f_1$ we have $P_5'(f_1)=f''(\xi_1)> 0$, so graph of the function $f$ is concave up at $\xi_1$. At quadruple zero $f_2$, $f\rightarrow f_2$ as $\xi \rightarrow \pm \infty$. Hence we may have a solitary wave solution with amplitude $f_1-f_2<0$. The other case gives unbounded solution.

\bigskip

\noindent \textbf{(5)\, One double and one simple zeros.}

\begin{figure}[!h]
\centering
\begin{center}
\includegraphics[angle=0,scale=.25]{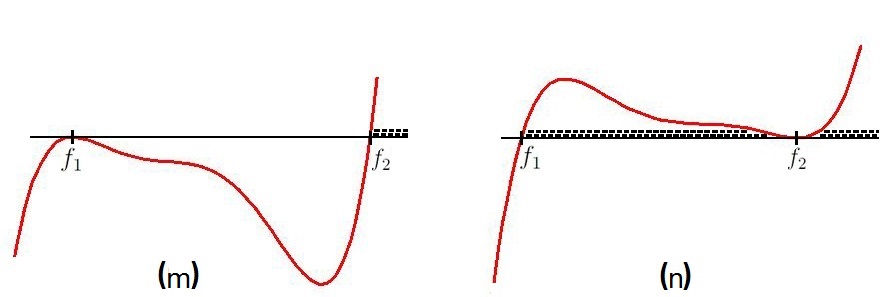}
\end{center}
\caption{Graphs of $P_5(f)$ having one double and one simple zeros}
\end{figure}

\noindent Consider the graph Figure 5.\textbf{(n)}. Here $f_1$ is a simple zero and $f_2$ is a double zero. The real solution occurs between $f_1$ and $f_2$. At $f_1$ we have $P_5'(f_1)=f''(\xi_1)> 0$, so graph of the function $f$ is concave up at $\xi_1$. At double zero $f_2$, $f\rightarrow f_2$ as $\xi \rightarrow \pm \infty$. Hence we have a solitary wave solution with amplitude $f_1-f_2<0$. The other case gives unbounded solution.

\bigskip

\noindent To sum up, we can give the following proposition for $\ell=3$ case.
\begin{pro}\label{proZEROanalysis}
Equation (\ref{P_5(f)}) may admit solitary wave solutions when the polynomial function $P_5(f)$ admits (i) one double and three simple zeros (ii) two double and one simple zeros (iii) one triple and two simple zeros (iv) one quadruple and one simple zeros (v) one double and one simple zeros.
\end{pro}

\section{Exact Solutions by Using the Chebyshev's Theorem}

\noindent The Chebyshev's Theorem is given as following \cite{Cheb}.

\begin{thm}\label{Chebthm} Let $a$, $b$, $c$, $\alpha$, $\beta$ be given real numbers and $\alpha\beta\neq 0$. The antiderivative
\begin{equation}\label{chebint}
I=\int x^a(\alpha +\beta x^b)^c dx
\end{equation}
is expressible by means of the elementary functions only in the three cases:
\begin{equation}\label{const1}
\displaystyle (1)\quad \frac{a+1}{b}+c \in \mathbb{Z},\quad (2)\quad \frac{a+1}{b} \in \mathbb{Z}, \quad (3)\quad c \in \mathbb{Z}.
\end{equation}
\end{thm}
The term $x^a(\alpha +\beta x^b)^c$ is called a differential binomial. Note that the differential binomial may be expressed in terms of the incomplete beta function and the hypergeometric function.
Let us define $u=\beta x^b/\alpha$. Then we have
\begin{eqnarray}\label{chebsoln}
\displaystyle I&=&  \frac{1}{b}\alpha^{\frac{a+1}{b}+c} \beta^{-\frac{a+1}{b}} B_y\Big(\frac{1+a}{b}, c-1  \Big).  \nonumber \\
&=& \frac{1}{1+a}\alpha^{\frac{a+1}{b}+c} \beta^{-\frac{a+1}{b}} u^{\frac{1+a}{b}}F\Big(\frac{a+1}{b}, 2-c; \frac{1+a+b}{b}; u \Big),
\end{eqnarray}
where $B_{y}$ is the incomplete beta function and $F(\tau, \kappa ; \eta ; u)$ is the hypergeometric function. Our aim is to transform the system (\ref{eqn1}) to $(y')^2=\bar{P}_{\ell+2}(y)$ by taking $f(\xi)=\gamma+\bar{\alpha}y(\bar{\beta}\xi)$. We can apply Chebyshev's Theorem to this equation if we assume that $\bar{P}_{\ell+2}(y)$ reduces to the form $\bar{P}_{\ell+2}(y)=A y^{-2a}(\alpha+\beta y^b)^{-2c}$, where $-2a-2c+b=\ell+2$. For $\ell=3$, let $u(x,t)=f(\xi)=\gamma+\bar{\alpha}y(\bar{\beta}\xi)$ in (\ref{P_5(f)}), then the equation becomes
\begin{equation}\label{poly}
(y')^2=\bar{P}_5(y)=\alpha_1 y^5+\alpha_2 y^4+\alpha_3 y^3+\alpha_4 y^2+\alpha_5 y+\alpha_6,
\end{equation}
where
\begin{eqnarray*}
\alpha_{1}&=&\frac{\bar{\alpha}^3}{2\bar{\beta}^2}, \\
\alpha_{2}&=&\frac{1}{\bar{\beta}^2}\Big(\frac{5\bar{\alpha}^2\gamma}{2}+3\bar{\alpha}^2c \Big), \\
\alpha_{3}&=&\frac{1}{\bar{\beta}^2}\Big(5\bar{\alpha}\gamma^2+6\bar{\alpha} c^2-2\bar{\alpha} d_1+12\bar{\alpha} c\gamma\Big), \\
\alpha_{4}&=&\frac{1}{\bar{\beta}^2}\Big(-4cd_1+4d_2+18c\gamma^2-6d_1\gamma+5\gamma^3+4c^3+18c^2\gamma \Big), \\
\alpha_{5}&=&\frac{1}{\bar{\alpha}\bar{\beta}^2}\Big(8d_3+12c\gamma^3+\frac{5\gamma^4}{2}+18c^2\gamma^2-8cd_1\gamma+8c^3\gamma+8d_2\gamma-6d_1\gamma^2\Big),\\
\alpha_{6}&=&\frac{1}{\bar{\alpha}^2\bar{\beta}^2}\Big(8d_3\gamma+6c^2\gamma^3+\frac{\gamma^5}{2}+8d_4-4cd_1\gamma^2-2d_1\gamma^3+4c^3\gamma^2+4d_2\gamma^2\Big).
\end{eqnarray*}
To apply the Chebyshev's Theorem \ref{Chebthm} we assume that $\bar{P}_5(y)$ reduces to the following form,
\begin{equation}
\bar{P_{5}}(y)=C y^{-2a}\,(\alpha+\beta y^{b})^{-2c},
\end{equation}
where $a$, $b$, $c$ are the constants in Theorem \ref{Chebthm}, $b-2a-2c=5$, and $C$ is a constant.

Here we present the cases mentioned in
the Proposition \ref{proZEROanalysis}. Other cases are given in Appendix A.

\vspace{0.5cm}

\begin{itemize}

\item[\textbf{1)}] Let $\bar{P}_5(y)=y(\alpha +\beta y^2)^2$. This form corresponds to the case of one simple or one simple and two double zeros. We have $y'=\pm y^{1/2}(\alpha+\beta y^2)$ so
\begin{equation}\label{caseA1}
\int y^{-1/2}(\alpha+\beta y^2)^{-1} dy=\pm \xi+A.
\end{equation}
\noindent Here $a=-1/2$, $b=2$, and $c=-1$. Hence
\begin{equation*}
\displaystyle (1)\quad \frac{a+1}{b}+c=-3/4 \notin \mathbb{Z},\quad (2)\quad \frac{a+1}{b}=1/4 \notin \mathbb{Z}, \quad (3)\quad c=-1 \in \mathbb{Z}.
\end{equation*}
\noindent For $(3)$, from (\ref{caseA1}), by letting $u=\beta y^b/\alpha$  we obtain
\begin{equation}
2\alpha^{-3/4}\beta^{-1/4}u^{1/4}F\Big(\frac{1}{4}, 3; \frac{5}{4}; u\Big)=\pm \xi+A.
\end{equation}
Here choose $\alpha=-1$, $\beta=1$, the integration constant $A=2$, plus sign in (\ref{caseA1}), we have
\begin{equation}
\mathrm{arctanh}(\sqrt{y})+\arctan(\sqrt{y})=-(\xi+2),
\end{equation}
and the graph of this solution for $\xi\leq -2$ is given in Figure 6
\begin{figure}[!h]
\centering
\begin{center}
\includegraphics[angle=0,scale=.30]{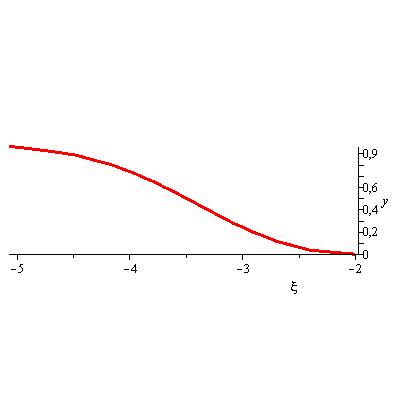}
\end{center}
\caption{Graph of a solution of $(y')^2=y(-1+y^2)^2$}
\end{figure}

\noindent This is a kink-type solution.
\item[\textbf{2)}] Let $\bar{P}_5(y)=y(\alpha +\beta y)^4$. This form corresponds to the case of one simple and one quadruple zeros. We have $y'=\pm y^{1/2}(\alpha+\beta y)^2$ so
\begin{equation}\label{caseA2}
\int y^{-1/2}(\alpha+\beta y)^{-2} dy=\pm \xi+A.
\end{equation}
\noindent Here $a=-1/2$, $b=1$, and $c=-2$. Hence\\
\begin{equation*}
\displaystyle (1)\quad \frac{a+1}{b}+c=-3/2 \notin \mathbb{Z},\quad (2)\quad \frac{a+1}{b}=1/2 \notin \mathbb{Z}, \quad (3)\quad c=-2 \in \mathbb{Z}.
\end{equation*}
\noindent For $(3)$, from (\ref{caseA2}), by letting $u=\beta y^b/\alpha$  we obtain
\begin{equation}
2\alpha^{-3/2}\beta^{-1/2}u^{1/2}F\Big(\frac{1}{2}, 4; \frac{3}{2}; u\Big)=\frac{\sqrt{y}}{\alpha(\beta y+\alpha)}+\frac{\arctan\Big(\frac{\beta \sqrt{y}}{\sqrt{\beta\alpha}} \Big) }{\alpha\sqrt{\beta\alpha}}= \pm \xi+A.
\end{equation}
Here choose $\alpha=1$, $\beta=1$, the integration constant $A=2$, plus sign in (\ref{caseA2}) we have
\begin{equation}
\displaystyle \frac{\sqrt{y}}{1+y}+\arctan(\sqrt{y})=\xi+2,
\end{equation}
and the graph of this solution for $-2\leq \xi<\pi/2-2$ is given in Figure 7
\begin{figure}[!h]
\centering
\begin{center}
\includegraphics[angle=0,scale=.30]{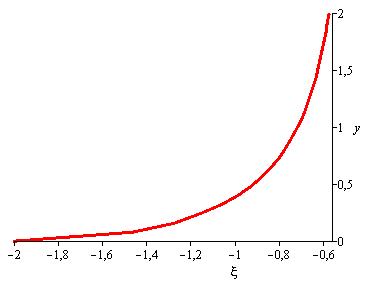}
\end{center}
\caption{Graph of a solution of $(y')^2=y(1+y)^4$}
\end{figure}

\bigskip
\item[\textbf{3)}] Let $\bar{P}_5(y)=y^2(\alpha +\beta y^3)$. This form corresponds to the case of one simple and one double zeros. We have $y'=\pm y(\alpha+\beta y^3)^{1/2}$ so
\begin{equation}\label{caseA3}
\int y^{-1}(\alpha+\beta y^3)^{-1/2} dy=\pm \xi+A.
\end{equation}
\noindent Here $a=-1$, $b=3$, and $c=-1/2$. Hence
\begin{equation*}
\displaystyle (1)\quad \frac{a+1}{b}+c=-1/2 \notin \mathbb{Z},\quad (2) \quad \frac{a+1}{b}=0 \in \mathbb{Z}, \quad (3)\quad c=-1/2 \notin \mathbb{Z}.
\end{equation*}
\noindent For $(2)$, from (\ref{caseA3}), we obtain
\begin{equation}
y=\Big(\frac{\alpha}{\beta}\Big(\tanh^2\Big(\frac{3}{2}\sqrt{\alpha}(A\pm \xi)\Big)-1\Big)\Big)^{1/3}.
\end{equation}
Here choose $\alpha=4$, $\beta=4$, the integration constant $A=2$, and  plus sign in (\ref{caseA3}) we obtain
\begin{equation}
y(\xi)=-(\mathrm{sech}(3\xi+6))^{2/3}
\end{equation}

\noindent and the graph of this solution is given in Figure 8

\begin{figure}[!h]
\centering
\begin{center}
\includegraphics[angle=0,scale=.30]{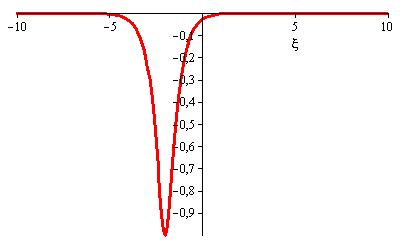}
\end{center}
\caption{Graph of a solution of $(y')^2=y^2(4+4y^3)$}
\end{figure}
\noindent This is clearly a solitary wave solution.

\item[\textbf{4)}] Let $\bar{P}_5(y)=y^2(\alpha +\beta y)^3$. This form corresponds to the case of one double and one triple zeros. We have $y'=\pm y(\alpha+\beta y)^{3/2}$ so
\begin{equation}\label{caseA4}
\int y^{-1}(\alpha+\beta y)^{-3/2} dy=\pm \xi+A.
\end{equation}
\noindent Here $a=-1$, $b=1$, and $c=-3/2$. Hence
\begin{equation*}
\displaystyle (1)\quad \frac{a+1}{b}+c=-3/2 \notin \mathbb{Z},\quad (2)\quad \frac{a+1}{b}=0 \in \mathbb{Z}, \quad (3)\quad c=-3/2 \notin \mathbb{Z}.
\end{equation*}
\noindent For $(2)$, from (\ref{caseA4}), we obtain
\begin{equation}
\frac{2}{\alpha\sqrt{\beta y+\alpha}}-\frac{2\mathrm{arctanh}\Big(\frac{\sqrt{\beta y+\alpha}}{\sqrt{\alpha}} \Big)}{\alpha^{3/2}} =\pm \xi+A.
\end{equation}
Here choose $\alpha=1$, $\beta=1$, $A=2$, and plus sign in (\ref{caseA4}) we have
\begin{equation}
\displaystyle \frac{2}{\sqrt{y+1}}-2\mathrm{arctanh}(\sqrt{y+1})=\xi+2,
\end{equation}
and the graph of this solution is given in Figure 9

\begin{figure}[!h]
\centering
\begin{center}
\includegraphics[angle=0,scale=.30]{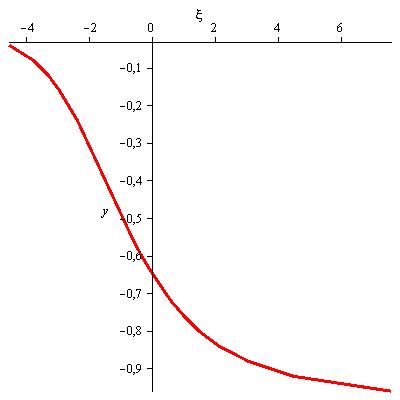}
\end{center}
\caption{Graph of a solution of $(y')^2=y^2(1+y)^3$}
\end{figure}

\bigskip
\item[\textbf{5)}] Let $\bar{P}_5(y)=y^4(\alpha +\beta y)$. This form corresponds to the case of one simple and one quadruple zeros. We have $y'=\pm y^2(\alpha+\beta y)^{1/2}$ so
\begin{equation}\label{caseA5}
\int y^{-2}(\alpha+\beta y)^{-1/2} dy=\pm \xi+A.
\end{equation}
\noindent Here $a=-2$, $b=1$, and $c=-1/2$. Hence
\begin{equation*}
\displaystyle (1)\quad \frac{a+1}{b}+c=-3/2 \notin \mathbb{Z},\quad (2)\quad \frac{a+1}{b}=-1 \in \mathbb{Z}, \quad (3)\quad c=-1/2 \notin \mathbb{Z}.
\end{equation*}
\noindent For $(2)$, from (\ref{caseA5}), by letting $u=\beta y^b/\alpha$ we obtain
\begin{equation}
- \alpha^{-3/2}\beta u^{-1}F\Big(-1, \frac{5}{2}; 0; u\Big) =-\frac{\sqrt{\beta y+\alpha}}{\alpha y}+\frac{\beta \mathrm{arctanh}\Big(\frac{\sqrt{\beta y+\alpha}}{\sqrt{\alpha}} \Big) }{\alpha^{3/2}}=\pm \xi+A.
\end{equation}

Take $\alpha=1$, $\beta=1$, $A=2$, and plus sign in (\ref{caseA5}) we get
\begin{equation}
\displaystyle \mathrm{arctanh}(\sqrt{y+1})-\frac{\sqrt{y+1}}{y}=\xi+2,
\end{equation}
and the graph of this solution for $\xi\geq -2$ is given in Figure 10

\begin{figure}[!h]
\centering
\begin{center}
\includegraphics[angle=0,scale=.30]{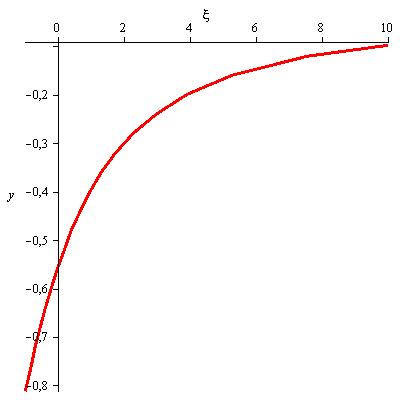}
\end{center}
\caption{Graph of a solution of $(y')^2=y^4(1+y)$}
\end{figure}

\end{itemize}

\section{An Alternative Method to Solve $(f')^2=P_n(f)$, When $n\geq 5$: Factorization of Polynomials}

\noindent When we have $(f')^2=P_n(f)$ for the reduced equation it becomes quite difficult to solve such equations
for $n\geq 5$. For this purpose, we shall introduce a new method which is based on the factorization of the polynomial $P_n(f)=P_{\ell+2}(f)$, $n\geq 5$.
Let the polynomial $P_{\ell+2}(f)$ have $\ell+2$ real roots i.e. $\displaystyle P_{\ell+2}(f)=B\prod_{i=1}^{\ell+2}(f-f_i)$, $B$ is a constant. Define a new function $\rho(\xi)$
so that $f=f(\rho(\xi))$. Hence we have $\displaystyle  (f')^2=\Big(\frac{df}{d\rho}\Big)^2\Big(\frac{d\rho}{d\xi}\Big)^2$. By taking
\begin{eqnarray*}
\displaystyle \Big(\frac{df}{d\rho}\Big)^2&=&\kappa \prod_{i=1}^{N_1}(f-f_i),\\
 \Big(\frac{d\rho}{d\xi}\Big)^2&=&\mu \prod_{i=1}^{N_2}(f-f_i),
\end{eqnarray*}
where $N_1+N_2=\ell+2$, $B=\kappa \mu$, we get a system of ordinary differential equations. Solving this system gives the solution of the degenerate coupled $\ell$-KdV equation.

For illustration, we start with
a differential equation where we know the solution. Consider
\begin{equation}
(y')^2=\alpha (y-y_1)(y-y_2)(y-y_3)(y-y_4).
\end{equation}
\noindent Let $y=y(\rho(\xi))$ so
\begin{equation*}
\displaystyle \Big(\frac{dy}{d\rho}\Big)^2\Big(\frac{d\rho}{d\xi}\Big)^2=\alpha (y-y_1)(y-y_2)(y-y_3)(y-y_4).
\end{equation*}
\noindent Take
\begin{eqnarray}
\Big(\frac{dy}{d\rho}\Big)^2&=&\kappa (y-y_1)(y-y_2)\label{exAlternative1}\\
\Big(\frac{d\rho}{d\xi}\Big)^2&=&\mu (y-y_3)(y-y_4)\label{exAlternative2},
\end{eqnarray}
where $\kappa\mu=\alpha$. We start with solving the equation (\ref{exAlternative1}). We have
\begin{equation*}
\displaystyle \frac{dy}{\sqrt{(y-y_1)(y-y_2)}}=\frac{dy}{b\sqrt{\Big(\frac{y-a}{b}\Big)^2-1}}=\pm\sqrt{\kappa}d\rho,
\end{equation*}
where $a=(y_1+y_2)/2$ and $b=(y_1-y_2)/2$. Let $(y-a)/b=\cosh z$. This gives us $dz=\pm\sqrt{\kappa}d\rho$. After taking the integral we obtain
\begin{equation*}
\mathrm{arccosh}\Big(\frac{y-a}{b}\Big)=\pm\sqrt{\kappa}\rho+C_1, \quad C_1\quad \mathrm{constant},
\end{equation*}
so that
\begin{equation}\label{solnEx}
y=h(\rho)=a+b\cosh(\sqrt{\kappa}\rho+C_2), \quad C_2\quad \mathrm{constant}.
\end{equation}
Now we insert this solution into Eq.(\ref{exAlternative2}). Let $\cosh(\sqrt{\kappa}\rho+C_2)=v$. Hence we get
\begin{equation}\label{exampleintegral}
\displaystyle \int \frac{dv}{\sqrt{\kappa}b\sqrt{v^2-1}\sqrt{(v-A)^2-B^2}}=\pm \sqrt{\mu}\xi+C_3,
\end{equation}
where $A=(y_3+y_4-y_1-y_2)/(y_1-y_2)$, $B=(y_3-y_4)/(y_1-y_2)$, and $C_3$ is a constant. Solving (\ref{exampleintegral}) and using (\ref{solnEx}) gives
\begin{equation}
\displaystyle y=\frac{y_1(y_3-y_2)+y_2(y_1-y_3)\mathrm{sn}^2((1/2)\sqrt{\kappa (y_3-y_2)(y_4-y_1)}(\sqrt{\mu}\xi+C_3),k)}{(y_3-y_2)+(y_1-y_3)\mathrm{sn}^2((1/2)\sqrt{\kappa (y_3-y_2)(y_4-y_1)}(\sqrt{\mu}\xi+C_3),k)},
\end{equation}
which was obtained in \cite{gur-pek}. Here $sn$ is the Jacobi elliptic function and $k$ is the elliptic modulus satisfying
$k^2=(f_2-f_4)(f_1-f_3)/(f_2-f_3)(f_1-f_4)$.

\noindent Now we apply our method to the case when the polynomial $P_n(f)$ is of degree $n=5$. We have several possible cases, but here we shall give the case when the
polynomial (\ref{P_5(f)}) has five real roots. Other cases are presented in Appendix B.

\noindent \textbf{Case 1.} If (\ref{P_5(f)}) has five real roots we can write it in the form
\begin{equation}\label{P_5(f)second}
(f')^2=\alpha(f-f_1)(f-f_2)(f-f_3)(f-f_4)(f-f_5),
\end{equation}
where $f_1, f_2, f_3, f_4,$ and $f_5$ are the zeros of the polynomial function $P_5(f)$.
Now define a new function $\rho(\xi)$ so that $f=f(\rho(\xi))$. Hence (\ref{P_5(f)second}) becomes
\begin{equation}
(f')^2=\Big(\frac{df}{d\rho}\Big)^2\Big(\frac{d\rho}{d\xi}\Big)^2=\alpha(f-f_1)(f-f_2)(f-f_3)(f-f_4)(f-f_5).
\end{equation}

\noindent \textbf{i.} Take
\begin{eqnarray}
\Big(\frac{df}{d\rho}\Big)^2&=&-(f-f_1)(f-f_2)(f-f_3)(f-f_4)\label{caseAnewmethod1}\\
\Big(\frac{d\rho}{d\xi}\Big)^2&=&-\alpha (f-f_5).\label{caseAnewmethod2}
\end{eqnarray}
\noindent In \cite{gur-pek} we have found the solutions of Eq.(\ref{caseAnewmethod1}). One of the solutions is
\begin{equation}\label{h(rho)caseA}
\displaystyle f=h(\rho)=\frac{f_4(f_3-f_1)+f_3(f_1-f_4)\mathrm{sn}^2((1/2)\sqrt{(f_1-f_3)(f_2-f_4)}\rho,k)}
{(f_3-f_1)+(f_1-f_4)\mathrm{sn}^2((1/2)\sqrt{(f_1-f_3)(f_2-f_4)}\rho,k)},
\end{equation}
where $k$ is the elliptic modulus satisfying $k^2=(f_3-f_2)(f_4-f_1)/(f_3-f_1)(f_4-f_2)$.  We use this solution in the equation (\ref{caseAnewmethod2}) and get
\begin{equation}\label{rhocaseA}
\displaystyle \int_0^{\rho} \frac{d\hat{\rho}}{\sqrt{f_5-h(\hat{\rho})}}=\int_0^{\rho}\sqrt{\frac{A+B\mathrm{sn}^2(\omega\hat{\rho},k)}{C+D\mathrm{sn}^2(\omega\hat{\rho},k)}}d\hat{\rho}  =\pm \sqrt{\alpha}\xi+C_1, \quad C_1 \quad \mathrm{constant},
\end{equation}
where
\begin{equation*}
A=f_3-f_1,\, B=f_1-f_4,\, C=(f_5-f_4)(f_3-f_1),\, D=(f_5-f_3)(f_1-f_4),
\end{equation*}
\noindent and $\omega=(1/2)\sqrt{(f_1-f_3)(f_2-f_4)}$. For particular values; $f_1=f_2=-1$, $f_3=3$, $f_4=0$, $f_5=2$, $\alpha=1$, $C_1=0$, and choosing plus sign in (\ref{rhocaseA}) we get the graph of the solution $f(\xi)$ given in Figure 11
\begin{figure}[!h]
\centering
\begin{center}
\includegraphics[angle=0,scale=.30]{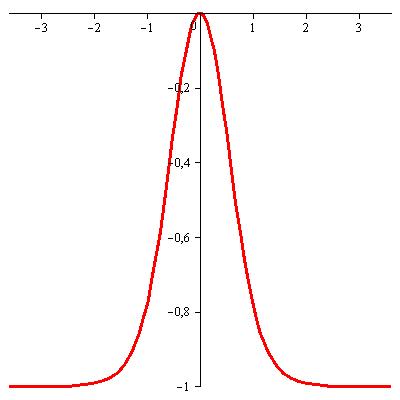}
\end{center}
\caption{Graph of the solution (\ref{h(rho)caseA}) with (\ref{rhocaseA}) for particular parameters}
\end{figure}

\noindent This is a solitary wave solution.
\bigskip

\noindent \textbf{ii.}Take
\begin{eqnarray}
\Big(\frac{df}{d\rho}\Big)^2&=&\kappa (f-f_1)(f-f_2)(f-f_3)\label{caseBnewmethod1}\\
\Big(\frac{d\rho}{d\xi}\Big)^2&=&\mu (f-f_4)(f-f_5),\label{caseBnewmethod2}
\end{eqnarray}
where $\kappa \mu=\alpha$. Eq.(\ref{caseBnewmethod1}) has a solution
\begin{equation}\label{h(rho)caseB}
f=h(\rho)=(f_2-f_1)\mathrm{sn}^2((1/2)\sqrt{f_3-f_1}(\sqrt{\kappa}\rho+C_1),k)+f_1,
\end{equation}
where $k$ satisfies $k^2=(f_1-f_2)/(f_1-f_3)$ and $C_1$ is a constant. We use this solution in Eq.(\ref{caseBnewmethod2}) and get
\begin{equation}\label{rhocaseB}
\displaystyle \int_0^{\rho} \frac{d\hat{\rho}}{\sqrt{(h(\hat{\rho})-f_4)(h(\hat{\rho})-f_5)}}=\int_0^{\rho} \frac{d\hat{\rho}}{\sqrt{A\mathrm{sn}^4(\omega\hat{\rho}+\delta,k)+B\mathrm{sn}^2(\omega\hat{\rho}+\delta,k)+C}} =\pm \sqrt{\mu}\xi+C_2,
\end{equation}
where $C_2$ is a constant and
\begin{equation*}
A=(f_2-f_1)^2,\, B=(f_2-f_1)(2f_1-f_4-f_5),\, C=(f_1-f_4)(f_1-f_5),\, \omega=(1/2)\sqrt{\kappa(f_3-f_1)},
\end{equation*}
and $\delta=(1/2)\sqrt{f_3-f_1}C_1$. For particular values; $f_1=1$, $f_2=2$, $f_3=3$, $f_4=-1$, $f_5=-2$, $\kappa=\mu=1$, $C_1=0$, and choosing plus sign in (\ref{rhocaseB}) we get the graph of the solution $f(\xi)$ given in Figure 12

\begin{figure}[!h]
\centering
\begin{center}
\includegraphics[angle=0,scale=.30]{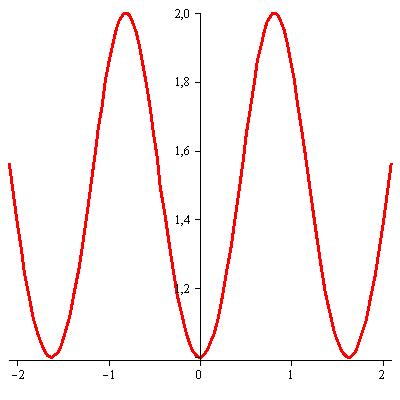}
\end{center}
\caption{Graph of the solution (\ref{h(rho)caseB}) with (\ref{rhocaseB})  for particular parameters}
\end{figure}
\smallskip
\noindent This solution is periodic.\\

\noindent \textbf{iii.} Take
\begin{eqnarray}
\Big(\frac{df}{d\rho}\Big)^2&=&\kappa (f-f_1)(f-f_2)\label{caseCnewmethod1}\\
\Big(\frac{d\rho}{d\xi}\Big)^2&=&\mu (f-f_3)(f-f_4)(f-f_5),\label{caseCnewmethod2}
\end{eqnarray}
where $\kappa \mu=\alpha$. Consider first the equation (\ref{caseCnewmethod1}).\\

\noindent We have
\begin{equation}
\displaystyle \frac{df}{\sqrt{(f-f_1)(f-f_2)}}=\pm \sqrt{\kappa} d\rho.
\end{equation}
This equality can be reduced to
\begin{equation}
\displaystyle \frac{df}{b\sqrt{\Big(\frac{f-a}{b}\Big)^2-1}}=\pm \sqrt{\kappa} d\rho,
\end{equation}
where $a=(f_1+f_2)/2$, $b=(f_1-f_2)/2$. By making the change of variables $(f-a)/b=u$, we get
\begin{equation}
\displaystyle \int \frac{du}{\sqrt{u^2-1}}=\pm \sqrt{\kappa}\rho +C_1,
\end{equation}
where $C_1$ is a constant. Thus we obtain
\begin{equation}
\displaystyle \mathrm{arc cosh} u=\mathrm{arc cosh} \Big(\frac{f-a}{b}\Big)=\pm \sqrt{\kappa}\rho +C_1,
\end{equation}
which yields
\begin{equation}\label{h(rho)caseC}
f=h(\rho)=a+b\cosh(\sqrt{\kappa}\rho +C_1), \quad a=(f_1+f_2)/2, \quad b=(f_1-f_2)/2.
\end{equation}
Now we use this result in (\ref{caseCnewmethod2}),
\begin{eqnarray}\label{rhocaseC}
\displaystyle &&\int_0^{\rho} \frac{d\hat{\rho}}{\sqrt{(h(\hat{\rho})-f_3)(h(\hat{\rho})-f_4)(h(\hat{\rho})-f_5)}}=
\nonumber\\&=&\int_0^{\rho}\frac{d\hat{\rho}}{\sqrt{A\cosh^3(\sqrt{\kappa}\hat{\rho} +C_1)
+B\cosh^2(\sqrt{\kappa}\hat{\rho} +C_1)+C\cosh(\sqrt{\kappa}\hat{\rho} +C_1)+D}}
\nonumber\\&=&\pm \sqrt{\mu} \xi+C_2,
\end{eqnarray}
\noindent where $C_2$ is a constant and
\begin{equation*}
A=b^3,\, B=b^2(3a-f_3-f_4-f_5),\, C=b\{(a-f_3)(a-f_4)+(a-f_3)(a-f_5)+(a-f_4)(a-f_5)\},
\end{equation*}
and $D=(a-f_3)(a-f_4)(a-f_5)$.

For particular values; $f_1=5$, $f_2=1$, $f_3=f_4=-3$, $f_5=-4$, $\kappa=4$, $\mu=1$, $C_1=C_2=0$, and choosing plus sign in (\ref{rhocaseC}), we get the graph of the solution $f(\xi)$ given in Figure 13
\begin{figure}[!h]
\centering
\begin{center}
\includegraphics[angle=0,scale=.30]{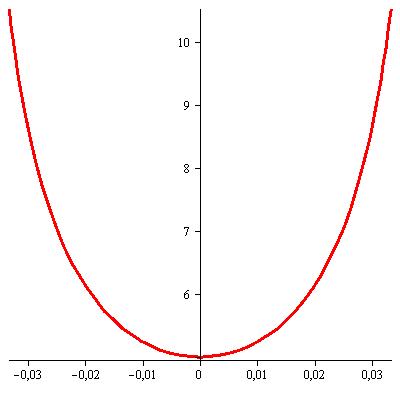}
\end{center}
\caption{Graph of the solution (\ref{h(rho)caseC}) for particular parameters}
\end{figure}

\bigskip

\noindent \textbf{iv.} Take
\begin{eqnarray}
\Big(\frac{df}{d\rho}\Big)^2&=&\kappa (f-f_1) \label{caseDnewmethod1}\\
\Big(\frac{d\rho}{d\xi}\Big)^2&=&\mu (f-f_2)(f-f_3)(f-f_4)(f-f_5),\label{caseDnewmethod2}
\end{eqnarray}
where $\kappa \mu=\alpha$. Consider the Eq.(\ref{caseDnewmethod1}). We have
\begin{equation}
\displaystyle \frac{df}{\sqrt{f-f_1}}=\pm \sqrt{\kappa} d\rho.
\end{equation}
Integrating both sides gives
\begin{equation}\label{h(rho)caseD}
\displaystyle f=h(\rho)=\Big(\pm \frac{\sqrt{\kappa}}{2}\rho+C_1\Big)^2+f_1, \quad C_1 \quad \mathrm{constant}.
\end{equation}
We use this result in the equation (\ref{caseDnewmethod2})
\begin{eqnarray}\label{rhocaseD}
 \displaystyle && \int_0^{\rho} \frac{d\hat{\rho}}{\sqrt{(h(\hat{\rho})-f_2)(h(\hat{\rho})-f_3)(h(\hat{\rho})-f_4)(h(\hat{\rho})-f_5)}}=
 \nonumber\\ &&=\int_0^{\rho} \frac{d\hat{\rho}}{\sqrt{(\frac{\kappa}{4}{\hat{\rho}}^2 \pm \sqrt{\kappa}C_1\hat{\rho}+A)(\frac{\kappa}{4}{\hat{\rho}}^2 \pm \sqrt{\kappa}C_1\hat{\rho}+B)(\frac{\kappa}{4}{\hat{\rho}}^2 \pm \sqrt{\kappa}C_1\hat{\rho}+C)(\frac{\kappa}{4}{\hat{\rho}}^2 \pm \sqrt{\kappa}C_1\hat{\rho}+D)}}\nonumber\\
&&=\pm \sqrt{\mu} \xi+C_2,
\end{eqnarray}
\noindent where $C_2$ is a constant and
\begin{equation*}
A=C_1^2+f_1-f_2,\, B=C_1^2+f_1-f_3,\, C=C_1^2+f_1-f_4,\, D=C_1^2+f_1-f_5.
\end{equation*}
For particular values, $f_1=1$, $f_2=f_3=2$, $f_4=3$, $f_5=4$, $\kappa=4$, $\mu=1$, and $C_1=C_2=0$, and choosing plus sign in (\ref{rhocaseD}) we get the graph of the solution $f(\xi)$ given in Figure 14
\begin{figure}[!h]
\centering
\begin{center}
\includegraphics[angle=0,scale=.30]{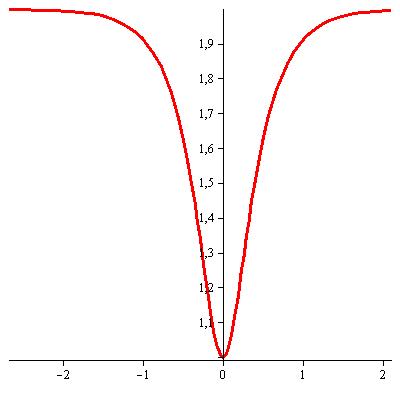}
\end{center}
\caption{Graph of the solution (\ref{h(rho)caseD}) for particular parameters}
\end{figure}
\noindent This is a solitary wave solution.

\section{Conclusion}
We study the degenerate $\ell$-coupled KdV equations. We reduce these equations into an ODE of the form $(f')^2=P_{\ell+2}(f)$, where
$P_{\ell+2}$ is a polynomial function of $f$ of degree $\ell+2$, $\ell\geq 3$. We give a general approach to
solve the degenerate $\ell$-coupled equations by introducing two new methods that one of them uses Chebyshev's Theorem and the other one is an alternative method, based on factorization of $P_{\ell+2}(f)$, $\ell\geq 3$. Particularly, for the degenerate three-coupled KdV equations we obtain solitary-wave, kink-type, periodic, or unbounded solutions by using these methods.

\section{Acknowledgment}
  This work is partially supported by the Scientific
and Technological Research Council of Turkey (T\"{U}B\.{I}TAK).\\

\noindent \textbf{APPENDIX A. Other Cases Using Chebyshev's Theorem for Degenerate Three-Coupled KdV Equation}

\smallskip

Here we present other cases that we use Cheyshev's theorem to solve degenerate three-coupled KdV equation.

\begin{itemize}
\item[\textbf{1)}] Let $\bar{P}_5(y)=y(\alpha+\beta y^4)$. This form corresponds to the case of one simple zero or three simple zeros. We have $y'=\pm y^{1/2}(\alpha+\beta y^4)^{1/2}$ so
\begin{equation*}
\int y^{-1/2}(\alpha+\beta y^4)^{-1/2} dy=\pm \xi+A.
\end{equation*}
\noindent Here $a=-1/2$, $b=4$, and $c=-1/2$. So we have
\begin{equation*}
\displaystyle (1)\, \frac{a+1}{b}+c=-3/8 \notin \mathbb{Z},\quad (2)\, \frac{a+1}{b}=1/8 \notin \mathbb{Z}, \quad (3)\, c=-1/2 \notin \mathbb{Z}.
\end{equation*}
\noindent Hence we cannot obtain a solution through the Chebyshev's theorem.

\item[\textbf{2)}] Let $\bar{P}_5(y)=y^3(\alpha +\beta y^2)$. This form corresponds to the case of one triple or two simple and one triple zeros. We have $y'=\pm y^{3/2}(\alpha+\beta y^2)^{1/2}$ so
\begin{equation*}
\int y^{-3/2}(\alpha+\beta y^2)^{-1/2} dy=\pm \xi+A.
\end{equation*}
\noindent Here $a=-3/2$, $b=2$, and $c=-1/2$. So we have
\begin{equation*}
\displaystyle (1)\quad \frac{a+1}{b}+c=-3/4 \notin \mathbb{Z},\quad (2)\quad \frac{a+1}{b}=-1/4 \notin \mathbb{Z}, \quad (3)\quad c=-1/2 \notin \mathbb{Z}.
\end{equation*}
\noindent Hence we cannot obtain a solution through the Chebyshev's theorem.

\item[\textbf{3)}] Let $\bar{P}_5(y)=y^3(\alpha +\beta y)^2$. This form corresponds to the case of one double and one triple zeros. We have $y'=\pm y^{3/2}(\alpha+\beta y)$ so
\begin{equation}\label{caseAp3}
\int y^{-3/2}(\alpha+\beta y)^{-1} dy=\pm \xi+A.
\end{equation}
\noindent Here $a=-3/2$, $b=1$, and $c=-1$. Hence
\begin{equation*}
\displaystyle (1)\quad \frac{a+1}{b}+c=-3/2 \notin \mathbb{Z},\quad (2)\quad \frac{a+1}{b}=-1/2 \notin \mathbb{Z}, \quad (3)\quad c=-1 \in \mathbb{Z}.
\end{equation*}
\noindent For $(3)$, from (\ref{caseAp3}), by letting $u=\beta y^b/\alpha$ we obtain
\begin{equation}
-2 \alpha^{-3/2}\beta^{1/2}u^{-1/2}F\Big(-\frac{1}{2}, 3; \frac{1}{2}; u\Big) =-\frac{2}{\alpha\sqrt{y}}+\frac{2\sqrt{\beta}\arctan \Big(\frac{\sqrt{\beta y}}{\sqrt{\alpha}} \Big)}{\alpha^{3/2}}=\pm \xi+A.
\end{equation}

\item[\textbf{4)}] Let $\bar{P}_5(y)=(\alpha +\beta y)^5$. This form corresponds to the case of a zero with multiplicity five. We have $y'=\pm (\alpha+\beta y)^{5/2}$ so
\begin{equation}\label{caseAp4}
\int (\alpha+\beta y)^{-5/2} dy=\pm \xi+A.
\end{equation}
\noindent Here $a=0$, $b=1$, and $c=-5/2$. Hence
\begin{equation*}
\displaystyle (1)\quad \frac{a+1}{b}+c=-3/2 \notin \mathbb{Z},\quad (2)\quad \frac{a+1}{b}=1 \in \mathbb{Z}, \quad (3)\quad c=-5/2 \notin \mathbb{Z}.
\end{equation*}
\noindent For $(2)$, from (\ref{caseAp4}), by letting $u=\beta y^b/\alpha$ we obtain
\begin{equation}
 \alpha^{-3/2}\beta^{-1} u F\Big(1, \frac{9}{2}; 2; u\Big) =-\frac{2}{3\beta(\beta y+\alpha)^{3/2}}=\pm \xi+A.
\end{equation}
\noindent Here we get
\begin{equation}
\displaystyle y=\frac{1}{\beta} \Big[\Big(\frac{4}{9\beta^2(\pm \xi+A)^2}\Big)-\alpha \Big].
\end{equation}

\item[\textbf{5)}] Let $\bar{P}_5(y)=(\alpha +\beta y^5)$. This form corresponds to the case of one simple zero. We have $y'=\pm (\alpha+\beta y^5)^{1/2}$ so
\begin{equation*}
\int (\alpha+\beta y^5)^{-1/2} dy=\pm \xi+A.
\end{equation*}
\noindent Here $a=0$, $b=5$, and $c=-1/2$. We have
\begin{equation*}
\displaystyle (1)\quad \frac{a+1}{b}+c=-3/10 \notin \mathbb{Z},\quad (2)\quad \frac{a+1}{b}=1/5 \notin \mathbb{Z}, \quad (3)\quad c=-1/2 \notin \mathbb{Z}.
\end{equation*}
\noindent Hence we cannot obtain a solution through the Chebyshev's theorem.

\end{itemize}

\bigskip

\noindent \textbf{APPENDIX B. Other Cases Using Alternative Method for Degenerate Three-Coupled KdV Equation}

\smallskip

Here we present the other cases for the method of factorization of the polynomial $P_5(f)$.

\noindent \textbf{Case 2.} If (\ref{P_5(f)}) has three real roots we can write it in the form
\begin{equation}\label{P_5(f)threerealroots}
(f')^2=\alpha(f-f_1)(f-f_2)(f-f_3)(af^2+bf+c),
\end{equation}
where $f_1$, $f_2$, and $f_3$ are the zeros of the polynomial function $P_5(f)$ of degree five and $b^2-4ac < 0$.
Now define a new function $\rho(\xi)$ so that $f=f(\rho(\xi))$. Hence (\ref{P_5(f)threerealroots}) becomes
\begin{equation}
(f')^2=\Big(\frac{df}{d\rho}\Big)^2\Big(\frac{d\rho}{d\xi}\Big)^2=\alpha(f-f_1)(f-f_2)(f-f_3)(af^2+bf+c),
\end{equation}

\noindent \textbf{i.} Take
\begin{eqnarray}
\Big(\frac{df}{d\rho}\Big)^2&=&\kappa (f-f_1)(f-f_2)(f-f_3)\label{case2inewmethod1}\\
\Big(\frac{d\rho}{d\xi}\Big)^2&=&\mu (af^2+bf+c),\label{case2inewmethod2}
\end{eqnarray}
where $\kappa \mu=\alpha$. Consider the first equation above. We have
\begin{equation}
\displaystyle \frac{df}{\sqrt{(f-f_1)(f-f_2)(f-f_3)}}=\pm \sqrt{\kappa} d\rho.
\end{equation}
We know from Case 1.ii that Eq.(\ref{case2inewmethod1}) has a solution
\begin{equation}
f=h(\rho)=(f_2-f_1)\mathrm{sn}^2((1/2)\sqrt{f_3-f_1}(\sqrt{\kappa}\rho+C_1))+f_1,
\end{equation}
where $C_1$ is a constant. We use this solution in (\ref{case2inewmethod2}) and we get
\begin{equation}\label{rhocase2i}
\displaystyle \int_0^{\rho} \frac{d\hat{\rho}}{\sqrt{ah^2(\hat{\rho})+b h(\hat{\rho})+c}}=\int_0^{\rho} \frac{d\hat{\rho}}{\sqrt{A\mathrm{sn}^4(\omega\hat{\rho}+\delta)+B\mathrm{sn}^2(\omega\hat{\rho}+\delta)+C}} =\pm \sqrt{\mu}\xi+C_2,
\end{equation}
where $C_2$ is a constant and
\begin{equation*}
A=a(f_2-f_1)^2,\, B=(2af_1+b)(f_2-f_1),\, C=(af_1^2+bf_1+c),\, \omega=(1/2)\sqrt{\kappa(f_3-f_1)},
\end{equation*}
and $\delta=(1/2)\sqrt{f_3-f_1}C_1$.

\bigskip

 \textbf{ii.} Take
\begin{eqnarray}
\Big(\frac{df}{d\rho}\Big)^2&=&\kappa (f-f_1)(f-f_2)\label{case2iinewmethod1}\\
\Big(\frac{d\rho}{d\xi}\Big)^2&=&\mu (f-f_3)(af^2+bf+c),\label{case2iinewmethod2}
\end{eqnarray}
where $\kappa \mu=\alpha$. Consider first the equation (\ref{case2iinewmethod1}). It has a solution
\begin{equation}\label{h(rho)case2ii}
f=h(\rho)=a_1+b_1\cosh(\sqrt{\kappa}\rho +C_1), \quad a_1=(f_1+f_2)/2, \quad b_1=(f_1-f_2)/2,
\end{equation}
as it is found in Case 1.iii. We use this solution in (\ref{case2iinewmethod2}) and get
\begin{eqnarray}\label{rhocase2ii}
\displaystyle &&\int_0^{\rho} \frac{d\hat{\rho}}{\sqrt{(h(\hat{\rho})-f_3)(ah^2(\hat{\rho})+b h(\hat{\rho})+c)}}=
\nonumber\\&=&\int_0^{\rho}\frac{d\hat{\rho}}{\sqrt{A\cosh^3(\sqrt{\kappa}\hat{\rho} +C_1)
+B\cosh^2(\sqrt{\kappa}\hat{\rho} +C_1)+C\cosh(\sqrt{\kappa}\hat{\rho} +C_1)+D}}
\nonumber\\&=&\pm \sqrt{\mu} \xi+C_2,
\end{eqnarray}
where $C_2$ is a constant and
\begin{equation*}
A=ab_1^3,\, B=b_1^2(3aa_1+b-f_3),\, C=b_1(3aa_1^2-2aa_1f_3+2ba_1-f_3b+c),
\end{equation*}
and $D=(a_1-f_3)(a_1^2a+ba_1+c)$.

\bigskip

\textbf{iii.} Take
\begin{eqnarray}
\Big(\frac{df}{d\rho}\Big)^2&=&\kappa (f-f_1)\label{case2iiinewmethod1}\\
\Big(\frac{d\rho}{d\xi}\Big)^2&=&\mu (f-f_2)(f-f_3)(af^2+bf+c),\label{case2iiinewmethod2}
\end{eqnarray}
where $\kappa \mu=\alpha$. Consider the equation (\ref{case2iiinewmethod1}). It has a solution
\begin{equation}\label{h(rho)case2iii}
\displaystyle f=h(\rho)=\Big(\pm \frac{\sqrt{\kappa}}{2}\rho+C_1\Big)^2+f_1, \quad C_1 \quad \mathrm{constant}
\end{equation}
as it is obtained in Case 1.iv. We use this result in Eq.(\ref{case2iiinewmethod2})
\begin{eqnarray}\label{rhocase2iii}
 \displaystyle && \int_0^{\rho} \frac{d\hat{\rho}}{\sqrt{(h(\hat{\rho})-f_2)(h(\hat{\rho})-f_3)(ah^2(\hat{\rho})+b h(\hat{\rho})+c)}}=
 \nonumber\\ &&=\int_0^{\rho} \frac{d\hat{\rho}}{\sqrt{(\frac{\kappa}{4}{\hat{\rho}}^2 \pm \sqrt{\kappa}C_1\hat{\rho}+A)(\frac{\kappa}{4}{\hat{\rho}}^2 \pm \sqrt{\kappa}C_1\hat{\rho}+B)(\frac{a\kappa^2}{16}{\hat{\rho}}^4\pm\frac{a\kappa^{3/2}}{2}C_1{\hat{\rho}}^3+C{\hat{\rho}}^2+D\hat{\rho}+E)}}\nonumber\\
&&=\pm \sqrt{\mu} \xi+C_2,
\end{eqnarray}
where $C_2$ is a constant and
\begin{equation*}
A=C_1^2+f_1-f_2,\, B=C_1^2+f_1-f_3,\, C=\kappa(b+6aC_1^2+2af_1)/4,
\end{equation*}
and
\begin{equation*}
D=\pm \sqrt{\kappa}C_1(2af_1+2aC_1^2+b),\, E=(f_1+C_1^2)(af_1+aC_1^2+b)+c.
\end{equation*}
\bigskip

\textbf{iv.} Take
\begin{eqnarray}
\Big(\frac{df}{d\rho}\Big)^2&=&\kappa (af^2+bf+c)\label{case2ivnewmethod1}\\
\Big(\frac{d\rho}{d\xi}\Big)^2&=&\mu (f-f_1)(f-f_2)(f-f_3),\label{case2ivnewmethod2}
\end{eqnarray}
where $\kappa \mu=\alpha$. Consider the first equation above. We have
\begin{equation*}
\displaystyle \int_0^f \frac{d\hat{f}}{\sqrt{a}\sqrt{\Big(\frac{\hat{f}+\frac{b}{2a}}{W}\Big)^2+1}}=\pm\sqrt{\kappa}\rho+C_1,
\end{equation*}
where $W=\sqrt{4ac-b^2}/2a$. Let $\Big(f+\frac{b}{2a}\Big)/W=\tan\theta$. Hence the above equality becomes
\begin{equation*}
\int_0^{\theta} \sec \hat{\theta} d\hat{\theta}=\sqrt{a}(\pm\sqrt{\kappa}\rho+C_1),
\end{equation*}
which gives
\begin{equation*}
\ln|\sec\theta+\tan\theta|=\sqrt{a}(\pm\sqrt{\kappa}\rho+C_1).
\end{equation*}
Hence after some simplifications and taking the integration constant zero we get
\begin{equation}\label{h(rho)case2iv}
\displaystyle f=h(\rho)=\pm \sqrt{4ac-b^2}\sinh(\sqrt{a\kappa}\rho)/2a.
\end{equation}
We use this result in (\ref{case2ivnewmethod2})
\begin{eqnarray}\label{rhocase2iv}
\displaystyle &&\int_0^{\rho} \frac{d\hat{\rho}}{\sqrt{(h(\hat{\rho})-f_1)(h(\hat{\rho})-f_2)(h(\hat{\rho})-f_3)}}=
\nonumber\\&=&\int_0^{\rho}\frac{d\hat{\rho}}{\sqrt{A\sinh^3(\sqrt{a\kappa}\hat{\rho})
+B\sinh^2(\sqrt{a\kappa}\hat{\rho})+C\sinh(\sqrt{a\kappa}\hat{\rho})+D}}
\nonumber\\&=&\pm \sqrt{\mu} \xi+C_2,
\end{eqnarray}
where $C_2$ is a constant and
\begin{equation*}
A=\pm (4ac-b^2)^{-3/2}/8a^3 , B=(b^2-4ac)(f_1+f_2+f_3)/4a^2,
\end{equation*}
and
\begin{equation*}
C=\pm \sqrt{4ac-b^2}(f_1f_3+f_2f_3+f_1f_2)/2a,\, D=-f_1f_2f_3.
\end{equation*}

\bigskip

\noindent \textbf{Case 3.} If (\ref{P_5(f)}) has just one real root we can write it in the form
\begin{equation}\label{P_5(f)onerealroot}
(f')^2=\alpha(f-f_1)(f^4+a_3f^3+a_2f^2+a_1f+a_0),
\end{equation}
where $f_1$ is the zero of the polynomial function $P_5(f)$ of degree five and the constants $a_i$, $i=0,1,2,3$ are so that $f^4+a_3f^3+a_2f^2+a_1f+a_0\neq 0$ for real $f$.

Now define a new function $\rho(\xi)$ so that $f=f(\rho(\xi))$. Hence (\ref{P_5(f)onerealroot}) becomes
\begin{equation}
(f')^2=\Big(\frac{df}{d\rho}\Big)^2\Big(\frac{d\rho}{d\xi}\Big)^2=\alpha(f-f_1)(f^4+a_3f^3+a_2f^2+a_1f+a_0).
\end{equation}

\noindent \textbf{i.} Take
\begin{eqnarray}
\Big(\frac{df}{d\rho}\Big)^2&=&\kappa (f-f_1)\label{case3inewmethod1}\\
\Big(\frac{d\rho}{d\xi}\Big)^2&=&\mu (f^4+a_3f^3+a_2f^2+a_1f+a_0),\label{case3inewmethod2}
\end{eqnarray}
where $\kappa \mu=\alpha$. Consider the first equation above. It has a solution
\begin{equation}\label{h(rho)case3i}
\displaystyle f=h(\rho)=\Big(\pm \frac{\sqrt{\kappa}}{2}\rho+C_1\Big)^2+f_1, \quad C_1 \quad \mathrm{constant}
\end{equation}
as it is obtained in Case 1.iv. We use this result in (\ref{case3inewmethod2}) and get
\begin{equation}\label{rhocase3i}
 \displaystyle \int_0^{\rho} \frac{d\hat{\rho}}{\sqrt{h^4(\hat{\rho})+a_3h^3(\hat{\rho})+a_2h^2(\hat{\rho})+a_1h(\hat{\rho})+a_0}}=\pm \sqrt{\mu}\xi+C_2,
\end{equation}
where $C_2$ is a constant. Here we do not get a simpler expression than the original equation. Hence it is meaningless
to use the method of factorization of the polynomial for this case.

\bigskip

\noindent \textbf{ii.} Take
\begin{eqnarray}
\Big(\frac{df}{d\rho}\Big)^2&=&\kappa (f^4+a_3f^3+a_2f^2+a_1f+a_0)\label{case3iinewmethod1}\\
\Big(\frac{d\rho}{d\xi}\Big)^2&=&\mu (f-f_1),\label{case3iinewmethod2}
\end{eqnarray}
where $\kappa \mu=\alpha$. Here for simplicity, we will use the form $f^4+bf^2+c$ instead of the polynomial function of degree four above. Since $\displaystyle f^4+bf^2+c=(f^2+b/2)^2-b^2/4+c$, we will take $c> b^2/4$ to get an irreducible polynomial. If $c=b^2/4$, then take $b>0$.
From $\Big(\frac{df}{d\rho}\Big)^2=\kappa (f^4+bf^2+c)$
\begin{equation*}
\displaystyle \int_0^{f} \frac{d \hat{f}}{\sqrt{\hat{f}^4+b\hat{f}^2+c}}=\pm \sqrt{\kappa}\rho+C_1 , \quad C_1 \quad \mathrm{constant}.
\end{equation*}
We have
\begin{equation*}
\displaystyle f(\rho)= \frac{\sqrt{2c}}{\sqrt{-b+\sqrt{-4c+b^2}}}\mathrm{sn}((\sqrt{2}/2) (C_1\pm \sqrt{\kappa}\rho ) \sqrt{-b+\sqrt{-4c+b^2}},k),
\end{equation*}
where $\displaystyle k=\frac{\sqrt{2}}{2} \sqrt{ \frac{-2c+b^2+b\sqrt{-4c+b^2}}{c}}$. Then we use this function $f$ in (\ref{case3iinewmethod2}) and get
\begin{equation}\label{rhocase3ii}
 \displaystyle \int_0^{\rho} \frac{\sqrt{A}d\hat{\rho}}{\sqrt{\sqrt{2c}\,\mathrm{sn}((A\sqrt{2}/2) (C_1\pm \sqrt{\kappa}\rho ),k)   -Af_1}}=\pm \sqrt{\mu}\xi+C_2, \quad C_2 \quad \mathrm{constant},
\end{equation}
where $A=\sqrt{-b+\sqrt{-4c+b^2}}$.

\bigskip

\noindent Now take $c=b^2/4$ with $b>0$. In this case we have
\begin{equation*}
\displaystyle \int_0^{f} \frac{d \hat{f}}{\hat{f}^2+b/2}=\pm \sqrt{\kappa}\rho+C_1 , \quad C_1 \quad \mathrm{constant},
\end{equation*}
which gives the solution
\begin{equation}
f=h(\rho)=\pm \frac{\sqrt{2b}}{2}\tan\Big( \frac{\sqrt{2b}}{2}(\sqrt{\kappa}\rho+\tilde{C}_1) \Big).
\end{equation}
Now use the equation (\ref{case3iinewmethod2}). We have
\begin{equation*}
\displaystyle \int_0^{\rho} \frac{d\hat{\rho}}{\sqrt{\pm \frac{\sqrt{2b}}{2}\tan\Big( \frac{\sqrt{2b}}{2}(\sqrt{\kappa}\hat{\rho}+\tilde{C}_1) \Big)-f_1 }}
=\pm \sqrt{\mu}\xi+C_2,
\end{equation*}
where $C_2$ is a constant. Solving the above equation gives
\begin{equation*}
\frac{\pm 4}{\sqrt{\kappa}b(A_2^2+A_3^2)A_3}\Big\{ \frac{\sqrt{2b}}{4}A_3^2\ln \Big[\frac{(A_1-A_2)^2+A_3^2}{(A_1+A_2)^2+A_3^2}\Big]
-b\arctan\Big(\frac{2A_1A_3}{A_3^2-A_1^2+A_2^2}\Big) \Big\}=\pm \sqrt{\mu}\xi+C_2,
\end{equation*}
where
\begin{equation*}
\displaystyle A_1=\sqrt{\pm 2\sqrt{2b}\tan \Big(\frac{\sqrt{2b}}{2}(\sqrt{\kappa}\rho+\tilde{C}_1)\Big)-4f_1}, A_2=\sqrt{\sqrt{4f_1^2+2b}-2f_1},
\end{equation*}
and $A_3=\sqrt{\sqrt{4f_1^2+2b}+2f_1}.$

\end{document}